\documentclass[12pt,english]{article}
\usepackage[latin9]{inputenc}
\usepackage{geometry}
\usepackage{color}
\usepackage{babel}
\usepackage{dsfont}
\usepackage{amsmath}
\usepackage{amsthm}
\usepackage{amssymb}
\usepackage{setspace}
\usepackage[authoryear,comma]{natbib}
\usepackage[pdfusetitle,
 bookmarks=true,bookmarksnumbered=false,bookmarksopen=false,
 breaklinks=false,pdfborder={0 0 1},backref=false,colorlinks=true]
 {hyperref}
\hypersetup{
 citecolor=blue,linkcolor=blue,urlcolor=blue}

\makeatletter

\providecommand{\tabularnewline}{\\}

\theoremstyle{plain}
    \ifx\thechapter\undefined
      \newtheorem{assumption}{\protect\assumptionname}
    \else
      \newtheorem{assumption}{\protect\assumptionname}[chapter]
    \fi
\theoremstyle{plain}
    \ifx\thechapter\undefined
	    \newtheorem{thm}{\protect\theoremname}
	  \else
      \newtheorem{thm}{\protect\theoremname}[chapter]
    \fi
\theoremstyle{plain}
    \ifx\thechapter\undefined
  \newtheorem{cor}{\protect\corollaryname}
\else
      \newtheorem{cor}{\protect\corollaryname}[chapter]
    \fi
\theoremstyle{plain}
    \ifx\thechapter\undefined
      \newtheorem{lem}{\protect\lemmaname}
    \else
      \newtheorem{lem}{\protect\lemmaname}[chapter]
    \fi

\usepackage{babel}
\usepackage{appendix}

\providecommand{\assumptionname}{Assumption}
\providecommand{\corollaryname}{Corollary}
\providecommand{\lemmaname}{Lemma}

\providecommand{\theoremname}{Theorem}

\makeatother

\providecommand{\assumptionname}{Assumption}
\providecommand{\corollaryname}{Corollary}
\providecommand{\lemmaname}{Lemma}
\providecommand{\theoremname}{Theorem}

\begin{document}
\title{Robust Inference with High-Dimensional Instruments}
\author{Qu Feng\thanks{Economics Division, Nanyang Technological University, Singapore, E-mail: \protect\protect\protect\href{http://qfeng@ntu.edu.sg}{qfeng@ntu.edu.sg}} \and Sombut Jaidee\thanks{Economics Division, Nanyang Technological University, Singapore, E-mail: \protect\protect\protect\href{http://sombut.jaidee@ntu.edu.sg}{sombut.jaidee@ntu.edu.sg}} \and Wenjie Wang\thanks{Economics Division, Nanyang Technological University, Singapore, E-mail: \protect\protect\href{http://wang.wj@ntu.edu.sg}{wang.wj@ntu.edu.sg}}}
\maketitle
\begin{abstract}

We propose a weak-identification-robust test for linear instrumental variable (IV) regressions with high-dimensional instruments, whose number is allowed to exceed the sample size.  
In addition, our test is robust to general error dependence, such as network dependence and spatial dependence. The test statistic takes a self-normalized form and the asymptotic validity of the test is established by using random matrix theory. Simulation studies are conducted to assess the numerical performance of the test, confirming good size control and satisfactory testing power across a range of various error dependence structures. 


\end{abstract}

\newpage

\section{Introduction\protect\label{sec:_Introduction}}

Various recent surveys in leading economics journals suggest that weak instruments remain important concerns for empirical practice. For instance, I.\cite{Andrews-Stock-Sun(2019)} survey 230 instrumental variable (IV) regressions from 17 papers published in the \textit{American Economic Review} (AER). They find that many of the first-stage F-statistics (and non-homoskedastic generalizations) are in a range that raises such concerns, and virtually all of these papers report at least one first-stage F with a value smaller than 10.
Similarly, in \citeauthor{lee2021}'s (\citeyear{lee2021}) survey of 123 AER articles involving IV regressions, 105 out of 847 specifications have first-stage Fs smaller than 10. 
Moreover, many IV applications involve a large number of instruments. 
For example, in their seminal paper, \cite{Angrist-Krueger(1991)} study the effect of schooling on wages by interacting three base instruments (dummies for the quarter of birth) with state and year of birth, resulting in 180 instruments. \cite{Hansen-Hausman-Newey(2008)} show that using the 180 instruments gives tighter confidence intervals than using the base instruments even after adjusting for the effect of many instruments. 
In addition, as pointed out by \cite{MS22},  in empirical papers that employ the ``judge design" (e.g., see \cite{maestas2013}, \cite{sampat2019}, and \cite{dobbie2018}), the number of instruments (the number of judges) is typically proportional to the sample size, and the famous Fama-MacBeth two-pass regression in empirical asset pricing (e.g., see \cite{fama1973}, \cite{shanken1992}, and \cite{anatolyev2022}) is equivalent to IV estimation with the number of instruments proportional to the number of assets. 
Similarly, \cite{belloni2012} consider an IV application involving more than one hundred instruments for the study of the effect of judicial eminent domain decisions on economic outcomes. 
\cite{carrasco2015} used many instruments in the estimation of the elasticity of intertemporal substitution in consumption.
Furthermore, as pointed out by \cite{Goldsmith(2020)}, the shift-share or Bartik 
instrument (e.g., see \cite{bartik1991} and \cite{blanchard1992}), which has been widely applied in many fields such as labor, public, development, macroeconomics, international trade, and finance, can be considered as a particular way of combining many instruments. For example, in the canonical setting of estimating the labor supply elasticity, the corresponding number of instruments is equal to the number of industries, which is also typically proportional to the sample size. 
Similar patterns of many IVs occur in wind-direction IVs, granular IVs,  local average treatment effect estimation, and Mendelian randomization.\footnote{E.g., see \cite{deryugina2019mortality}, \cite{bondy2020crime}, \cite{gabaix2024}, \cite{blandhol2022}, \cite{boot2024}, \cite{sloczynski2024should}, \cite{davey2003}, and  \cite{davies2015}.}

The problem of inference under many instruments has been well studied in the literature (e.g., see the literature review below). However, to our knowledge, there still does not exist a valid inference method that is robust to the high dimensionality of instruments and general error dependence, such as network and spatial dependence, simultaneously. To fill this gap, we propose in this paper a novel procedure for inference under high-dimensional IVs. Our test statistic takes a self-normalized form, and we establish the asymptotic validity of our test under general error dependence structure, by using random matrix theory (RMT). Additionally, we derive the power property of our test under many instruments. Monte Carlo simulations are conducted to assess the numerical performance of the test, confirming good size control and satisfactory testing power across a range of various error dependence structures. 
\vspace{0.1in} 

\noindent
\textbf{Literature Review.} The contributions in the present paper is related to the large literature on many (weak) instruments.\footnote{See, for example, \cite{Kunitomo1980}, \cite{morimune1983}, \cite{Bekker(1994)}, \cite{donald2001}, \cite{chamberlain2004}, \cite{Chao-Swanson(2005)}, \cite{stock2005}, \cite{han2006}, \cite{Andrews-Stock(2007)}, \cite{Hansen-Hausman-Newey(2008)}, \cite{Newey-Windmeijer(2009)}, \cite{anderson2010}, \cite{kuersteiner2010}, \cite{anatolyev2011}, \cite{belloni2011}, \cite{okui2011}, \cite{belloni2012}, \cite{carrasco2012}, 
\cite{Chao(2012)}, \cite{Haus2012},  
\cite{K13},
\cite{hansen2014}, \cite{carrasco2015}, \cite{Wang_Kaffo_2016}, \cite{kolesar2018}, \cite{EK18}, \cite{solvsten2020}, \cite{CNT23}, \cite{boot2024},
among others.} 
In the context of many instruments and heteroskedasticity, \cite{Chao(2012)} and \cite{Haus2012} provide standard errors for Wald-type inferences that are based on JIVE and jackknifed versions of the limited information maximum likelihood (LIML) and \citeauthor{Fuller(1977)}'s (\citeyear{Fuller(1977)}) estimators (HLIM and HFUL). These estimators are more robust to many instruments than the commonly used two-stage least squares (TSLS) estimator because they can correct the bias caused by the high dimension of IVs.\footnote{Specifically, the rate of growth of the concentration parameter, which measure the overall instrument strength, is denoted as $\mu_n^2$. JIVE, HLIM, and HFUL remain consistent with heteroskedastic errors even when instrument weakness is such that $\mu_n^2$ is slower than the number of instruments $K$, provided that $\mu_n^2/\sqrt{K} \rightarrow \infty$ as the number of observations $n \rightarrow \infty$ (\citealp{Chao(2012), Haus2012}). In contrast, TSLS is less robust to instrument weakness as it is shown to be consistent only under homoskedasticity if $\mu_n^2/K \rightarrow \infty$ \citep{Chao-Swanson(2005)}. In simulations derived from the data in \cite{Angrist-Krueger(1991)}, which is representative of empirical labor studies with many instrument concerns, \citet[Section IV]{Angrist-Frandsen2022} show that such bias-corrected estimators outperform the TSLS that is based on the instruments selected by the least absolute shrinkage and selection operator (LASSO) introduced in \cite{belloni2012} or the random forest-fitted first stage introduced in \cite{athey2019}.}  
Furthermore, \cite{carrasco2012}, \cite{carrasco2015, carrasco2016efficient}, \cite{hansen2014}, and \cite{carrasco2017} proposed regularization approaches for two-stage least squares, limited information maximum likelihood, and jackknife IV (\citep{Angrist(1999)}) estimators. Under many weak moment asymptotics, \cite{Newey-Windmeijer(2009)} provide new variance estimators for the jackknife GMM and the class of generalized empirical likelihood (GEL) estimators, which includes the continuous updating estimator (CUE) and EL estimator as special cases.\footnote{In the linear heteroskedastic IV model, consistency and asymptotic normality of CUE require $m^2/n \rightarrow 0$ and $m^3/n \rightarrow 0$, respectively, where $m$ and $n$ denote the number of moment conditions and the sample size (e.g., see p.689 of \cite{Newey-Windmeijer(2009)}). Such conditions are needed to simultaneously control the estimation error for all the elements of the heteroskedasticity consistent weighting matrix. Somewhat stronger rate conditions are required for other GEL estimators.}

However, the Wald-type inference methods are invalid under weak identification, which occurs when the concentration parameter remains bounded as the sample size increases to infinity. In this case,  all the estimators mentioned earlier become inconsistent, and there is no consistent test for the structural parameter of interest.\footnote{E.g., see Section 3 of \cite{MS22}.}  
For weak-identification-robust inference based on the classical AR test, \cite{Andrews-Stock(2007)} show its validity under many instruments, but their IV model is homoskedastic and requires the number of instruments to diverge slower than the cube root of the sample size ($K^3/n \rightarrow 0$). \cite{Newey-Windmeijer(2009)} proposed a GMM-AR test under many (weak) moment conditions but imposed the same rate condition on $K$. \cite{anatolyev2011} constructed a modified AR test that allows the number of instruments to be proportional to the sample size but requires homoskedastic errors, and \cite{Kaffo-Wang(2017)} proposed a bootstrap version of \cite{anatolyev2011}'s test. Furthermore, \cite{carrasco2016} first proposed a ridge-regularized AR test that allows for $K$ being larger than $n$ with homoskedastic errors. \cite{Bun-Farbmacher-Poldermans(2020)} compared the centered and uncentered GMM-AR test and identified a missing degrees-of-freedom correction when $K/n \rightarrow 0$. 

Recently, \cite{crudu2021} and \cite{MS22} proposed jackknifed versions of the AR test in a model with many instruments and general heteroskedasticity. 
\cite{DKM24} developed a ridge-regularized version of the jackknife AR test. 
\citet[Section 2.3]{tuvaandorj2024} established the validity of a permutation AR test under heteroskedasticity and diverging $K$, requiring $K^3/n \rightarrow 0$. \cite{boot-ligtenberg(2023)} developed a dimension-robust AR test based on continuous updating but relied on an invariance assumption.  
Furthermore, \cite{LWZ2024dimension} proposed a dimension-agnostic bootstrap AR test by deriving strong approximation results for both test statistic and its bootstrap version. Their bootstrap AR test is valid with heteroskedastic errors uniformly across a broad asymptotic regime for $K$, spanning from fixed to diverging faster than the sample size.  
In addition to the AR tests, \cite{Matsushita2020} propose a jackknife LM test. \cite{Lim2024} consider a linear combination of jackknife AR, jackknife LM, and orthogonalized jackknife LM tests and find that the resulting conditional linear combination (CLC) test has good power properties in a variety of scenarios. 
\cite{N23} proposed a dimension-robust version of \cite{Kleibergen(2002)}'s K test, and his method relies on a sparse $\ell_1$-regularized estimation of $\rho(Z_i)$, the conditional correlation between the endogenous variable and the outcome error. 
\cite{Yap24} proposed a leave-three-out variance estimator, in the same spirit as that proposed by \cite{AS23}, for inference under heterogeneous treatment effects.
The above methods are robust to weak identification, many instruments, and heteroskedastic errors. 
However, the literature on inference procedures that are robust to more general error dependence structure remains sparse. \cite{ligtenberg2023} proposed a leave-cluster-out AR test that is robust to cluster dependence structure under many weak instruments. \nocite{Wang-Zhang2024}
\vspace{0.1in}

The rest of this paper is organized as follows. In Section \ref{sec:_Model_and_stat}, we formally define the observed intermediary, which is used to formulate our proposed statistic. Then, we establish their limiting distributions with power theories. 
Section \ref{sec:Simulation} reports the finite sample performance of our proposed statistics across multiple scenarios through Monte Carlo simulations. 
The Appendices are provided separately as the supplementary document, including 
proof of theorems and relevant lemmas.

\section{Model and Test Statistic \protect\label{sec:_Model_and_stat}}

We consider the linear IV regression with a scalar outcome $y_i$, a scalar endogenous variable $x_i$, and a $K \times 1$ vector of instruments $z_i$ such that 
\begin{align}\label{eq: model}
y_i = x_i \beta + \varepsilon_i, \quad  x_i = z_i'\pi + v_i, \quad \forall i \in [N].
\end{align}
Following the literature on many instruments, we assume that there are no (low-dimensional) controls included in our model as they can be partialled out from $(Y_i,X_i,Z_i)$. 
The source of endogeneity is caused by a correlation between $\varepsilon_{i}$ and $v_{i}$, which represent the errors associated with the structural and the first-stage equation, respectively. The two sequences of regression errors, $\left\{ \varepsilon_{i}\right\} _{i=1}^{N}$ and $\left\{ v_{i}\right\} _{i=1}^{N}$, follow a mean-zero random process whose functional forms are unknown and potentially heteroskedastic. The unknown parameters are $\beta\in\mathbb{R}$ and $\pi\in\mathbb{R}^{K\times1}$, where $K$ is comparable or exceeds its sample size. 

We are interested in testing the null hypothesis 
\begin{equation}
H_{0}:\beta=\beta_{0}\text{ versus }H_{1}:\beta\neq\beta_{0}\label{eq:_null_hypothesis},
\end{equation}
even when the identification for the structural parameter of interest $\beta$ may be weak.
We focus on the model with a scalar endogenous variable for two reasons. First, in many empirical applications of IV regressions, there is only one endogenous variable (as can be seen from the surveys by \cite{Andrews-Stock-Sun(2019)} and \cite{lee2021}). Second, the asymptotic results derived in the paper extend directly to the general case of full-vector inference with multiple endogenous variables. 
Additionally, for the subvector inference, one may use a projection approach \citep{Dufour-Taamouti(2005)} after implementing our test on the whole vector of endogenous variables.\footnote{Alternative subvector inference methods for IV regressions (e.g., see \cite{GKMC(2012)}, \cite{Andrews(2017)}, and \cite{GKM(2019), GKM(2021)}) provide a power improvement over the projection approach under fixed $K$. However, whether they can be applied to the current setting is unclear.  
Also, \cite{Wang-Doko(2018)} and \cite{Wang(2020)} show that bootstrap tests based on the standard subvector AR statistic may not be robust to weak identification even under fixed $K$ and conditional homoskedasticity.}


To proceed, let us denote $y^*_{i}=y_{i}-x_{i}\beta_{0}.$ 
Let a singular value decomposition of the normalized instrument matrix be expressed as 
\begin{align}
\frac{Z}{\sqrt{N}}:=\sum_{\ell=1}^{\min\left\{ K,N\right\} }\sqrt{\lambda_{\ell}}q_{\ell}w_{\ell}^{'},
\end{align}
where $Z:=\left(z_{1},\ldots,z_{N}\right)^{'}\in\mathbb{R}^{N\times K},$ $\lambda_{\ell}$ denotes the $\ell$-th largest non-zero eigenvalue of the matrix $S_{N}:=\frac{1}{N}Z^{'}Z,$ $q_{\ell}$ and $w_{\ell}$ denotes the associated orthonormal eigenvector of the $\ell$-th largest eigenvalue of $\underline{S}_{N}:=\frac{1}{N}ZZ^{'}$ and $S_{N}$, respectively. 

To conduct the hypothesis testing in \eqref{eq:_null_hypothesis}, the associated test statistic proposed in this paper can be obtained by equating the sample analogue of
\begin{equation}
\mathbb{E}\left[\lambda_{\ell}\cdot\left(\frac{1}{\gamma_{j}^{2}}\left(q_{\ell}^{'}Y^{*}\right)^{2}-1\right)\right]=0,\label{eq:_a_moment_of_interest}
\end{equation}
where $Y^{*}=Y-X\beta_{0}$, $Y=\left(y_{1},\ldots,y_{N}\right)^{'}$, 
$X=\left(x_{1},\ldots,x_{N}\right)^{'}$, 
and $\gamma_{j}^{2}:=\mathbb{E}\varepsilon_{j}^{2}$. 
Similar (high-dimensional) moment condition was studied in \citet{Feng2024} to overcome the curse of dimensionality of coefficients in linear regression models with a presence of (unconditional) heteroskedasticity and autocorrelation of an unknown form. 
After substituting $\gamma_{j}^{2}$ with its plug-in estimator under the null, namely $\frac{Y^{*'}Y^{*}}{N}$, the statistic of interest is obtained by evaluating its associated sample counterpart: 
\begin{align}
M_{N} & =\frac{1}{N}\sum_{\ell=1}^{\min\left\{K,N\right\} }\lambda_{\ell}\left(N\left(q_{\ell}^{'}\underline{Y}\right)^{2}-1\right)\notag \\
 & =\sum_{\ell=1}^{\min\left\{ K,N\right\} }\left(q_{\ell}^{'}\underline{Y}\right)^{2}\lambda_{\ell}-\frac{1}{N}\sum_{\ell=1}^{\min\left\{ K,N\right\} }\lambda_{\ell},
\end{align}
where $\underline{Y}:=Y^{*}/\left(Y^{*'}Y^{*}\right)^{1/2}\in\left(0,1\right)$. Then, with a proper convergence rate (shown in Theorem \ref{thm:_oracle_null}), our (oracle) statistic of interest is given by
\begin{align}
Q_{N} & :=\sqrt{\frac{N^{2}}{2\text{tr}\left(\Sigma^{2}\right)}}\left(\sum_{\ell=1}^{\min\left\{ K,N\right\} }\left(q_{\ell}^{'}\underline{Y}\right)^{2}\lambda_{\ell}-\frac{1}{N}\sum_{\ell=1}^{\min\left\{ K,N\right\} }\lambda_{\ell}\right) \notag \\
 & =\sqrt{\frac{N^{2}}{2\text{tr}\left(\Sigma^{2}\right)}}\left(\underline{Y}{}^{'}\underline{S}_{N}\underline{Y}-\frac{1}{N}\text{tr}\left(\underline{S}_{N}\right)\right).
\end{align}
Unless stated otherwise, we denote $\underline{a}:a/\left(a^{'}a\right)^{1/2}$ a self-normalized version of any column vector $a.$  

To build intuition, suppose for the moment that the errors are homoskedastic, i.e., $\mathbb{E}(\varepsilon \varepsilon' |Z) = \gamma^2I_N$, where $\varepsilon = (\varepsilon_1, ..., \varepsilon_N)'$. 
By the law of iterated expectation,  
\begin{align}
    \mathbb{E}\left[\lambda_{\ell}\cdot\left(\frac{1}{\gamma^{2}}\left(q_{\ell}^{'}Y^{*}\right)^{2}-1\right)\right] 
    & = \mathbb{E} \left[ \lambda_{\ell} \frac{1}{\gamma^2} \left(
    q'_{\ell} \mathbb{E}(\varepsilon \varepsilon'|Z) q_{\ell}\right) -\lambda_{\ell} \right] \notag \\
    &= \mathbb{E} \left[ \lambda_{\ell} \frac{1}{\gamma^2} \left(\gamma^2 \cdot q'_{\ell} I_N q_{\ell} \right) - \lambda_{\ell} \right] =0,
\end{align}
under the null hypothesis, where $\ell$ is randomly chosen from the set $\{1, ..., \min\{K,N\}\}$.
We note that $q'_{\ell}q_{s}$ and $w'_{\ell}w_s$ are equal to one when $\ell = s$ for all $\ell, s = 1, ..., \min\{K,N\}$ but zero otherwise. On the other hand, under the alternative $\beta \neq \beta_0$, 
\begin{align}
\mathbb{E}\left[\lambda_{\ell}\cdot\left(\frac{1}{\gamma^{2}}\left(q_{\ell}^{'}Y^{*}\right)^{2}-1\right)\right] 
= \frac{1}{\gamma^2} \mathbb{E} \left[ \lambda_{\ell} \cdot (q'_{\ell} X \Delta)^2 \right] > 0, 
\end{align}
where $\Delta = \beta - \beta_0$.

For general error dependence, i.e., $\mathbb{E}(\varepsilon \varepsilon' | Z) = \Psi$, it is expected that under the null hypothesis, 
\begin{align}
    \mathbb{E}\left[\lambda_{\ell}\cdot\left(\frac{1}{\Psi_{ii}}\left(q_{\ell}^{'}Y^{*}\right)^{2}-1\right)\right] 
    = 
    \mathbb{E} \left[ \lambda_{\ell} \left( \frac{q'_{\ell}\varepsilon}{\Psi_{ii}}\right)^2 - \lambda_{\ell} \right] \rightarrow 0, 
\end{align}
where $\Psi^2_{ii}$ denotes the $i$-th diagonal entry of $\Psi$. The proposed test becomes robust against general error dependence with high-dimensional IVs if there exists, for some $i \in \{1, ..., N\}$, the sequence $\{ (q_{\ell}' \varepsilon \cdot \Psi_{ii}^{-1})^2\}_{\ell=1}^{\min\{K, N\}}$
converges to one in probability as $K$ and $N$ grow, 
where the quantity $q_{\ell}'\varepsilon \cdot \Psi_{ii}^{-1}$ is the inner product of an eigen basis generated by the Gram matrix $\underline{S}_N$ and the normalized regression errors. 
We establish the asymptotic validity of our inference procedure under the following regularity conditions. 

\begin{assumption}[High-dimensional asymptotic regime]
 The dimension of instruments is considered high-dimension such that $K\rightarrow\infty,$ $N\rightarrow\infty$, and $K/N \rightarrow c\in\left(0,\infty\right)$. \label{ass:_std_highdim_regime} 
\end{assumption}
With rank deficiency generally observed in sample covariance matrices, this regime is widely regarded as a standard high-dimensional paradigm in the literature \citep[see e.g.,][]{Ledoit2002, Bai2007, Onatski2013}. This paper introduces a more adaptable framework at a general level of intensity beyond that of \citet{Khan2024}, where the number of covariates may exceed its sample size.

\begin{assumption}[Instruments]
 The data generating process of $Z=\left(z_{1},\ldots,z_{N}\right)^{'}\in\mathbb{R}^{N\times K}$ is given by $z_{i}=\Sigma^{1/2}f_{i},$ where $f_{i}=\left(f_{1i},\ldots,f_{Ki}\right)^{'}$ for all $i=1,\ldots,N$, $\Sigma^{1/2}$ is the square root of a deterministic non-negative definite Hermitian matrix $\Sigma,$ and $\lambda_{max}\left(\Sigma\right)=O\left(1\right)$. Moreover, $f_{i}$ consists of $K$ independent and identically distributed (i.i.d.) random variables with mean zero, unit variance, and finite fourth moment. \label{ass:_instruments} 
\end{assumption}

Assumption \ref{ass:_instruments} is recognized as a typical design in RMT literature \citep[e.g.,][]{Karoui2008, dobriban2018high, hastie2022surprises, Li2024, farbmacher2024revisiting, Zhang2025},which we employ to characterize correlations within instruments. The instruments, therefore, possess a weak form of cross-sectional dependence defined in \citet{Chudik2011}. Qualitative instruments can also be included, but require standardization or variable transformation to ensure zero mean, unit variance, and a finite fourth moment (see, Theorem \ref{thm:_oracle_null}). 
Assumption \ref{ass:_instruments} allows us to apply RMT, which helps to generalize the error dependence structure, in the current setting with high-dimensional instruments, as can be seen in Assumption \ref{ass:_errors_iid_X} below. 

\begin{assumption}
Denote $a_{-i}$ a whole sequence of the random variable $a$ with the $i$-th entry removed. The two unobserved errors in the IV model are given by
\[
\varepsilon_{i}=\Xi_{\varepsilon}\left(i,\varepsilon_{-i}^{'},v^{'},\mathcal{M}^{'}\right), \;\; \text{and} \;\;
v_{i}=\Xi_{v}\left(i,v_{-i}^{'},\varepsilon^{'},\mathcal{M}^{'}\right),
\]
where $\Xi_{\varepsilon}$ and $\Xi_{v}$ are unknown functions and $\mathcal{M}=\left(m_{1}^{'},\ldots,m_{N}^{'}\right)^{'}$ where $m_{i}$ denotes a vector of exogenous variables that jointly explains the variation of the error at individual $i$ for all $i=0, \ldots, N$. 
The data generating process of $z$ is independent of $\varepsilon$ and $v$. \label{ass:_errors_iid_X}
\end{assumption}
The permitted structure of errors are general and flexible, accommodating complicated network or spatial dependence structure into account. Assumption \ref{ass:_errors_iid_X} ensures the mean independence of $\varepsilon$ and $v$ against instruments $z$. Specifically, $\mathbb{E}\left(\left.f_{i}^{\delta}\right|\varepsilon_{i},v_{i}\right)=\mathbb{E}\left(f_{i}^{\delta}\right)$, for $0<\delta\leq4$. 

Define the self-normalized vector $u:=\left(u_{1},\ldots,u_{N}\right)$ where $u_{i}=\varepsilon_{i}/\left(\varepsilon^{'}\varepsilon\right)^{1/2}.$ The following theorem provides the null limiting distribution of $Q_{N}$, the oracle statistic, under high-dimensional IVs with a presence of (unconditional) heteroskedasticity in error processes.
\begin{thm}[Oracle statistic]
 \label{thm:_oracle_null} Suppose that Assumptions \ref{ass:_std_highdim_regime}-\ref{ass:_errors_iid_X} are satisfied. Further suppose that (i) $\mathbb{E}\left(f_{11}^{4}\right)=\kappa<\infty$,and (ii) $\sum_{i=1}^{N}\left|u_{i}^{3}\right|=o_{\mathbb{P}}\left(1\right).$ 
 Then, under the null hypothesis in \eqref{eq:_null_hypothesis}, 
\[
Q_{N}=\sqrt{\frac{N^{2}}{2\text{tr}\left(\Sigma^{2}\right)}}\left(\underline{Y}{}^{'}\underline{S}_{N}\underline{Y}-\frac{1}{N}\text{tr}\left(\underline{S}_{N}\right)\right)\overset{d}{\rightarrow}\mathcal{N}\left(0,1\right).
\]
\end{thm}
Replacing $\text{tr}\left(\Sigma^{2}\right)$ with the consistent estimator proposed by \citet{Li2012}, 
we have the following asymptotic result for the feasible statistic.
\begin{cor}[Feasible statistic]
Under Assumptions \ref{ass:_std_highdim_regime}-\ref{ass:_errors_iid_X}, and all conditions in Theorem \ref{thm:_oracle_null}, we further assume that $\sum_{i=1}^{N}u_{i} = O_{\mathbb{P}}\left(1\right).$ Then, \label{cor:_Feasible_stat}
\[
\widehat{Q}_{N} := \sqrt{\frac{N^{2}}{2\widehat{\text{tr}\left(\Sigma^{2}\right)}}}\left(\underline{Y}^{'}\underline{S}_{N}\underline{Y}-\frac{1}{N}\text{tr}\left(\underline{S}_{N}\right)\right)\overset{d}{\rightarrow}\mathcal{N}\left(0,1\right),
\]
where $\widehat{\text{tr}\left(\Sigma^{2}\right)}:=\frac{1}{N\left(N-1\right)}\sum_{i\neq j}^{N}\left(z_{i}^{'}z_{j}\right)^{2}$ is a ratio-consistent estimator of $\text{tr}\left(\Sigma^{2}\right)$. 
\end{cor}

Let $\Delta = \beta - \beta_0$. The alternative with local departure is defined as
\begin{equation}
H_{1}:h^{2}:=\frac{N}{\left(2\text{tr}\left(\Sigma^{2}\right)\right)^{2/5}}\Delta^{2}>0,\label{eq:=000020local=000020alternative}
\end{equation}
where $h = \Delta \cdot N^{1/2}/ (2 tr(\Sigma^2))^{1/5}$ denotes a (deterministic) level of departure. 
By construction, the quantity $\left(\frac{\left(2\text{tr}\left(\Sigma^{2}\right)\right)^{2/5}}{N}\right)^{1/2}$ implies a rate of signal strength in the coefficient of interest that the test can detect. 


\begin{thm}[Power theory]
 \label{thm:_power_theory} Let Assumptions \eqref{ass:_std_highdim_regime}-\eqref{ass:_errors_iid_X} hold. In addition, suppose that: (i) $\frac{1}{N}\varepsilon^{'}\varepsilon\overset{p}{\rightarrow}\alpha_{1}$ and $\frac{1}{N}v^{'}v\overset{p}{\rightarrow}\alpha_{2}$ where $\alpha_{1},\alpha_{2}\in\left(0,\infty\right)$, (ii) $\mathbb{E}\left(f_{11}^{8}\right)<\infty$, 
 and (iii) $\pi'\pi/\sqrt{K}$ is bounded. 
 Then, $Q_{N}-\varpi_{N}\overset{d}{\rightarrow}\mathcal{N}\left(0,1\right),$ where 
\begin{align*}
\varpi_{N} & =\frac{N}{\left\Vert \varepsilon\right\Vert ^{2}\left(2\text{tr}\left(\Sigma^{2}\right)\right)^{1/10}}h^{2} \pi'\left(S_{N}^{2} 
-\frac{1}{N}\text{tr}\left(S_{N}\right)I_{N}\right)\pi \\
& + \frac{N}{\left\Vert \varepsilon\right\Vert ^{2}\left(2\text{tr}\left(\Sigma^{2}\right)\right)^{1/10}}h^{2} 
v'\left(\frac{1}{N} \underline{S}_N - \frac{1}{N^2} tr(S_N) I_N \right)v \\
 & + \frac{2 N^{1/2}}{\left\Vert \varepsilon\right\Vert ^{2}\left(2\text{tr}\left(\Sigma^{2}\right)\right)^{3/10}}hv^{'}\left(\underline{S}_{N} - \frac{1}{N}\text{tr}\left(S_{N}\right)I_{N}\right)\varepsilon.
\end{align*}
\end{thm}

Several comments are in order. 
First, the power of the feasible statistic is analogous to that of the oracle one, since one merely replaces the unknown quantity $\text{tr}\left(\Sigma^{2}\right)$ with its consistent estimator. Namely $\left(\widehat{Q}_{N}-Q_{N}\right)+Q_{N}-\varpi_{N}=Q_{N}-\varpi_{N}+o_{\mathbb{P}}\left(1\right).$ 
Second, Condition (ii) can be relaxed to a finite fourth moment with a cost of technical complication.

\section{Monte Carlo Simulations }

\label{sec:Simulation}

In this part, we present the power performance of our proposed statistics across a range of data generating processes. The statistical software adopted for numerical results is MATLAB 2023a with the default seed. We further note that all parameters and functional forms postulated in the error processes are arbitrarily chosen. To design regimes of high dimensionality, we fix the sample size to $400$, then vary the number of instruments so that the data intensities ($K/n$) are $1/4,1/2,1,2$ and $3$. The numerical size (power) of statistics is calculated as the probability of null rejections over 1,000 replications, given the data is generated under the null (alternative) hypothesis. Throughout the analysis, an intercept is set to $2$ and a level of significance is set to 0.05. 
\begin{table}
\caption{The table illustrates finite sample performance of the feasible test statistic $\widehat{Q}_N$. The number of observations $N$ is fixed at 400 associated with $K/N$ of $1/4,1/2,1,2$ and $3.$ $h$, the local deviation from the true value, is set to $0,1,2$ and $5$, where zero indicates the empirical size of the test. Three data generating processes of errors are considered with the degree of endogeneity $\rho$ being $.5,.9$ and $-.9$ respectively.}

\centering{}%
\begin{tabular}{ccccccccccccc}
\\
\\
 &  & \multicolumn{3}{c}{{[}NET-E{]}} &  & \multicolumn{3}{c}{{[}SAR-E{]}} &  & \multicolumn{3}{c}{{[}MUL-E{]}}\tabularnewline
$\left(K/N,\rho\right)$ &  & .5 & .9 & $-.9$ &  & .5 & .9 & $-.9$ &  & .5 & .9 & $-.9$\tabularnewline
\hline 
\\
$\left[h=0\right]$ &  &  &  &  &  &  &  &  &  &  &  & \tabularnewline
1/4 &  & 6.0 & 6.6 & 6.9 &  & 5.3 & 6.9 & 7.2 &  & 6.8 & 5.6 & 6.9\tabularnewline
1/2 &  & 6.1 & 6.0 & 5.1 &  & 6.8 & 6.9 & 6.0 &  & 6.3 & 5.2 & 6.1\tabularnewline
1 &  & 6.7 & 5.9 & 4.8 &  & 7.0 & 5.8 & 6.1 &  & 4.6 & 5.8 & 6.2\tabularnewline
2 &  & 6.1 & 6.0 & 6.3 &  & 6.3 & 4.7 & 5.2 &  & 4.7 & 5.4 & 4.8\tabularnewline
3 &  & 6.5 & 5.9 & 7.7 &  & 5.1 & 5.8 & 6.2 &  & 6.7 & 6.6 & 5.0\tabularnewline
$\left[h=1\right]$ &  &  &  &  &  &  &  &  &  &  &  & \tabularnewline
1/4 &  & 41.0 & 35.5 & 59.9 &  & 38.3 & 34.1 & 63.1 &  & 37.9 & 34.1 & 62.9\tabularnewline
1/2 &  & 32.1 & 29.5 & 59.3 &  & 33.9 & 29.6 & 62.5 &  & 35.5 & 28.0 & 58.9\tabularnewline
1 &  & 29.7 & 25.9 & 57.1 &  & 32.6 & 26.3 & 58.3 &  & 27.7 & 25.8 & 58.5\tabularnewline
2 &  & 25.3 & 22.2 & 58.3 &  & 24.0 & 19.5 & 57.7 &  & 24.2 & 22.2 & 56.8\tabularnewline
3 &  & 21.9 & 19.6 & 55.3 &  & 20.2 & 20.6 & 56.7 &  & 24.0 & 21.0 & 59.7\tabularnewline
$\left[h=2\right]$ &  &  &  &  &  &  &  &  &  &  &  & \tabularnewline
1/4 &  & 88.4 & 81.5 & 100 &  & 86.6 & 81.0 & 99.8 &  & 89.5 & 83.0 & 99.9\tabularnewline
1/2 &  & 86.6 & 76.7 & 100 &  & 86.0 & 77.0 & 100 &  & 83.2 & 77.1 & 100\tabularnewline
1 &  & 79.7 & 71.4 & 100 &  & 78.2 & 67.8 & 100 &  & 79.7 & 69.2 & 100\tabularnewline
2 &  & 71.5 & 61.9 & 100 &  & 71.0 & 56.4 & 100 &  & 71.8 & 60.0 & 100\tabularnewline
3 &  & 63.3 & 50.7 & 100 &  & 62.9 & 54.9 & 100 &  & 62.2 & 53.7 & 100\tabularnewline
$\left[h=5\right]$ &  &  &  &  &  &  &  &  &  &  &  & \tabularnewline
1/4 &  & 99.9 & 100 & 100 &  & 100 & 99.7 & 100 &  & 99.8 & 99.4 & 100\tabularnewline
1/2 &  & 100 & 99.8 & 100 &  & 100 & 99.5 & 100 &  & 100 & 99.8 & 100\tabularnewline
1 &  & 99.9 & 99.3 & 100 &  & 100 & 98.8 & 100 &  & 99.8 & 99.3 & 100\tabularnewline
2 &  & 99.4 & 98.0 & 100 &  & 99.6 & 97.6 & 100 &  & 99.3 & 97.5 & 100\tabularnewline
3 &  & 98.1 & 95.5 & 100 &  & 98.5 & 95.2 & 100 &  & 98.1 & 93.2 & 100\tabularnewline
\\
\hline
\end{tabular}
\end{table}

The errors in the structural equation, i.e.,  $\varepsilon$, are generated by one of the three following processes: (i) the network dependent process, (ii) the spatial dependent process, and (iii) the multiplicative heteroskedasticity of a mixture of two i.i.d. processes. 
Specifically, for the network dependence process, we follow the simulation design of the network formation model in Section 5 of \cite{Kojevnikov2021}, and set $\gamma=0.5$ in (5.1) of their model. The processes (ii) and (iii) were also implemented in \citet{Feng2024}. For the case with spatial dependence, the errors follow the spatially autoregressive error process of order one studied by \cite{kim2011spatial}, that is, 
\begin{align*}
    \varepsilon = \rho_{s} W e,
\end{align*}
where $\rho_s=0.8$ and the innovation $e$ is independently generated by a standardized $t$-distribution with 5 degrees of freedom. In addition, $W=[c_iw_{ij}]$ is a $N \times N$ spatial weighting matrix. To specify such a matrix, we set all diagonal entries to zero and quantify the off-diagonal entry $w_{ij}$ to one if and only if $w_{ij}=a>0.5$, where $a$ is drawn from a uniform distribution over an interval $[0,1]$, otherwise zero. Denote $c_i$ a scalar that standardizes each row of $W$ to sum to unity. 
For the multiplicative heteroskedasticity and autocorrelation of a mixture of two i.i.d. processes, we let 
\begin{align*}
    \varepsilon_i = s_1 \left[\zeta_i (\omega_i + 2.4) -s_2 \right],
\end{align*}
where $\zeta_i = 1 + \left( a i/N \right)$
and $\omega_i = \sqrt{1-0.7^2} \eta_{1,i} + 0.7 \eta_{2,i}$,
$\eta_{1,i}$ is generated by i.i.d. standardized $t$-distribution with 5 degrees of freedom, and $\eta_{2,i}$ is generated by i.i.d. standardized chi-square distribution with 6 degrees of freedom. As a result, the underlying process $\omega_i$ is considered both heavy-tailed and highly-skewed. 
The scalar $s_1:= 289a^2/300 + (1+ a/2)^2$ and $s_2:= 1.2(a+2)$ normalizes the random variable $\varepsilon_i$ to mean zero and variance one. Under this design, the errors are equi-correlated with a correlation approximately equal to $0.52$ for $i, j=1, ..., N$. The impact of the skedastic component is set to $a=10$.

For brevity, the three processes are abbreviated as {[}NET-E{]}, {[}SPA-E{]}, and {[}MUL-E{]}, respectively. The errors in the first-stage equation are correlated with the other given by $v_{i}=\rho\varepsilon_{i}+\sqrt{1-\rho^{2}}\eta_{3,i}$, where $\eta_3$ is independently generated by a standardized $t$-distribution with $5$ degrees of freedom (abbreviated as $t_{5}$ hereafter). The two processes are strongly correlated, taking values of $\rho\in\left\{ -.9,.5,.9\right\} .$ 

Being consistent with our asymptotic theory, the instruments are generated according to a latent factor model studied in Section 2.1 of \citet{bhattacharya2011sparse}. Namely, 
\begin{align*}
    z_{i}=\Lambda\eta_{4,i}+\Sigma^{1/2}f_{i},
\end{align*}
where $\eta_{4,i}\in\mathbb{R}^{3}$ and $f_{i}\in\mathbb{R}^{K}$ are independently drawn from $t_{5}$ and the standard normal distribution, respectively. $\Lambda$ is a $K\times3$ loading matrix such that $\Lambda^{'}\Lambda=\text{diag}\left(6,5,3\right)$. The population covariance matrix $\Sigma$ is the Toeplitz matrix with the $\left(i,j\right)$-th elements equaling to $0.7^{\left|i-j\right|}$. The strength of IVs is set to be weak such that $\pi^{'}\pi=1$ throughout the simulations. 

The finite sample performance of the feasible statistic is analogous to that of the oracle one. To save space, we thus only report the results for the feasible test. We highlight several findings below.
First, the simulation results illustrate that the empirical size of our test is close to the nominal level across different settings of the number of instruments ($K/N$), the degree of endogeneity ($\rho$), and the error dependence structure.
Second, we note that the power of our test improves as the departure $h$ increases and attains unity, ensuring complete distributional separation between hypotheses, when $h$ exceeds 5. 
Third, in line with our asymptotic theory, the correlation between the two error processes influences the power of the test. A negative contemporary dependence enhances the power overall, as well as its magnitude. 

\section{Conclusion}

In this paper, we propose a weak-identification-robust test for linear instrumental variable (IV) regressions with high-dimensional instruments, whose number is allowed to exceed the sample size.  
In addition, our test is robust to general error dependence, such as network dependence and spatial dependence. The test statistic takes a self-normalized form and the asymptotic validity of the test is established by using RMT. Simulation studies are conducted to assess the numerical performance of the test, confirming good size control and satisfactory testing power across a range of various error dependence structures. For future research agenda, we note that although our new test is robust to high-dimensional instruments under general error dependence, it is not robust to the case with a fixed number of instruments. 
On the other hand, \cite{LWZ2024dimension} recently proposed a dimension-agnostic bootstrap AR test under many weak instruments and (independent) heteroskedastic errors by using strong approximation. For dimension-robust inference under the current setting, it may be interesting to explore a bootstrap-based inference procedure and combine it with RMT. 



\newpage{}

\bibliographystyle{chicago}

\appendix

\section{Technical details}

Let $\mathbb{E}_{n}\left(\cdot\right)$ denote the conditional expectation given the sigma-algebra $\mathcal{F}_{N,n}$ generated by the sequence $\left\{ \varepsilon_{1},\ldots,\varepsilon_{N},v_{1},\ldots,v_{n},f_{1},\ldots f_{n}\right\} $, i.e., $\mathcal{F}_{N,n}=\sigma\left(\varepsilon_{1},\ldots,\varepsilon_{N},v_{1},\ldots,v_{n},f_{1},\ldots,f_{n}\right)$. And, $\mathbb{E}\left(\cdot\right)$ denotes the unconditional expectation with respect to the random variables $f$, $v$ and $\varepsilon$. We define $e_{j}$ as a $N$-dimensional vector whose entries are all zero except for the $j$-th entry, which is equal to one.

By definition, $Z^{'}=\Sigma^{1/2}F$ where $F=\left(f_{1},\ldots,f_{N}\right),$ and each entry in the $p$-dimensional vector $f_{n}$, unless otherwise stated, follows a standardized independent and identically distributed (iid) process with finite fourth moment. 
Recall that $\underline{S}_{N}:=\frac{1}{N}ZZ^{'}=\frac{1}{N}F^{'}\Sigma F$ by construction. We use $\lambda_{i}\left(A\right)$ to denote the $i$-th eigenvalue of a non-negative definite square matrix $A$. Let $M$ denote a universal constant, which may take different values depending on the context. Unless stated otherwise, we denote $\underline{a}:a/\left(a^{'}a\right)^{1/2}$ a self-normalized of any column vector $a.$ The following lemmas are stated for the sake of convenience. 
\begin{lem}[Lemma 2.7 in \citealp{bai1998apx}]
For $X^{'}=\left(X_{1},\ldots,X_{p}\right)$ i.i.d. standardized entries and $C$ $p\times p$ real-valued matrix, we have, for any $p\geq2,$\label{lem:=000020Concentration=000020inequality} 
\[
\mathbb{E}\left|X^{'}CX-\text{tr}\left(C\right)\right|^{p}\leq K\left[\left(\mathbb{E}\left|X_{1}\right|^{4}\text{tr}CC^{'}\right)^{p/2}+\mathbb{E}\left|X_{1}\right|^{2p}\text{tr}\left(CC^{'}\right)^{p/2}\right].
\]
\end{lem}

\subsection{Proof of Theorem \ref{thm:_oracle_null}}

The proof is identical to Theorem 1 in \citep{Feng2024} so we omitted the proof.

\subsection{Proof of Corollary \ref{cor:_Feasible_stat}}

The proof is identical to Corollary 1 in \citep{Feng2024} so we omitted the proof.

\subsection{Proof of Theorem \ref{thm:_power_theory}}

Under the alternative \eqref{eq:=000020local=000020alternative}, we have $Y=X\Delta+\varepsilon$. The test statistic $Q_{N}$ can be decomposed to 
\begin{align*}
Q_{N} & =\sqrt{\frac{N^{2}}{2\text{tr}\left(\Sigma^{2}\right)}}\left(\frac{\left(X\Delta + \varepsilon\right)^{'}}{\left\Vert X\Delta + \varepsilon\right\Vert }\left(\underline{S}_{N}\right)\frac{\left(X\Delta + \varepsilon\right)}{\left\Vert X\Delta + \varepsilon\right\Vert }-\frac{1}{N}\text{tr}\left(\underline{S}_{N}\right)\frac{\left\Vert X\Delta + \varepsilon\right\Vert ^{2}}{\left\Vert X\Delta + \varepsilon\right\Vert ^{2}}\right)\frac{\frac{1}{\left\Vert \varepsilon\right\Vert ^{2}}}{\frac{1}{\left\Vert \varepsilon\right\Vert ^{2}}}\\
 & =\frac{\sqrt{\frac{N^{2}}{2\text{tr}\left(\Sigma^{2}\right)}}\left(u^{'}\underline{S}_{N}u-\frac{1}{N}\text{tr}\left(\underline{S}_{N}\right)\left(\frac{\left\Vert X\Delta + \varepsilon\right\Vert ^{2}}{\left\Vert \varepsilon\right\Vert ^{2}}\right)\right)+\frac{\Delta^{2}X^{'}ZZ^{'}X}{\sqrt{2\text{tr}\left(\Sigma^{2}\right)}\left\Vert \varepsilon\right\Vert ^{2}}+\frac{2\Delta X^{'}ZZ^{'}\varepsilon}{\sqrt{2\text{tr}\left(\Sigma^{2}\right)}\left\Vert \varepsilon\right\Vert ^{2}}}{\frac{\left\Vert X\Delta+\varepsilon\right\Vert ^{2}}{\left\Vert \varepsilon\right\Vert ^{2}}}\\
 & =:\frac{J_{1}}{J_{2}},
\end{align*}
where $\left\Vert X\Delta + \varepsilon\right\Vert ^{2} = \Delta^{2}X^{'}X + 2 \Delta X^{'}\varepsilon + \varepsilon^{'}\varepsilon$. We aim to show the following:

(i) $J_{1}\overset{d}{\rightarrow}\mathcal{N}\left(\varpi_{N},1\right),$ where $\varpi_{N}$ is defined in Theorem \ref{thm:_power_theory},
and

(ii) $J_{2}=\frac{\Delta^{2}X^{'}X + 2\Delta X^{'}\varepsilon+\varepsilon^{'}\varepsilon}{\left\Vert \varepsilon\right\Vert ^{2}}\overset{p}{\rightarrow}1.$ 
\begin{proof}[Proof of the statement (ii)]
 First consider the denominator $J_{2}$, it suffices to show that $J_{21}:=\frac{\Delta^{2}X^{'}X}{\left\Vert \varepsilon\right\Vert ^{2}}$ and $J_{22}:=\frac{2\Delta^{}X^{'}\varepsilon}{\left\Vert \varepsilon\right\Vert ^{2}}$ converge in probability to zero. 
 Recall that the signal length of $\Delta$ is directly proportional to $\left(\frac{\text{tr}^{2/5}\left(\Sigma^{2}\right)}{N}\right)^{1/2}$. 
 In addition, the first-stage equation in matrix form is given by $X=Z\pi+v.$ Then, 
\begin{align*}
0\leq\frac{\Delta^{2}\left(Z\pi+v\right)^{'}\left(Z\pi+v\right)}{\left\Vert \varepsilon\right\Vert ^{2}} & =\frac{\Delta^{2}\pi^{'}Z^{'}Z\pi+2\Delta^{2}v^{'}Z\pi+\Delta^{2}v^{'}v}{\left\Vert \varepsilon\right\Vert ^{2}}\\
 & =:J_{211}+J_{212}+J_{213}.
\end{align*}

To show $J_{21}\overset{p}{\rightarrow}0$, it suffices to show that $J_{21i}\overset{p}{\rightarrow}0$ for $i=1,2,3$. By Condition (i), $J_{213}=\frac{\Delta^{2}\frac{1}{N}v^{'}v}{\frac{1}{N}\left\Vert \varepsilon\right\Vert ^{2}}\overset{p}{\rightarrow}0.$ By the trace inequality, 
\begin{align*}
J_{211} & =\frac{\Delta^{2}\pi^{'}S_{N}\pi}{\frac{1}{N}\left\Vert \varepsilon\right\Vert ^{2}}\leq\frac{h^{2}\text{tr}^{2/5}\left(\Sigma^{2}\right)}{N}\frac{\lambda_{\max}\left(S_{N}\right)\pi^{'}\pi}{\frac{1}{N}\left\Vert \varepsilon\right\Vert ^{2}}\\
 & = O_{\mathbb{P}}\left(\frac{h^{2}K^{2/5}\pi^{'}\pi}{N}\right)\overset{p}{\rightarrow}0.
\end{align*}
By Jensen's inequality, and the trace inequality, 
\begin{align}
\mathbb{E}\left|v^{'}Z\pi\right| & \leq\left(\mathbb{E}\left(v^{'}Z\pi\pi^{'}Zv\right)\right)^{1/2}\nonumber \\
 & =\left(v^{'}\mathbb{E}\left(\left.\sum_{i=1}^{N}\sum_{j=1}^{N}e_{i}f_{i}^{'}\Sigma^{1/2}\pi\pi^{'}\Sigma^{1/2}f_{j}e_{j}^{'}\right|v\right)v\right)^{1/2}\nonumber \\
 & =\left(\mathbb{E}\left(\pi^{'}\Sigma^{1/2}\mathbb{E}\left(\left.\sum_{i=1}^{N}\sum_{j=1}^{N}f_{j}f_{i}^{'}\right|v\right)\Sigma^{1/2}\pi v_{i}v_{j}\right)\right)^{1/2}\nonumber \\
 & =\left(\pi^{'}\Sigma\pi\cdot\mathbb{E}\left(\sum_{i=1}^{N}v_{i}^{2}\right)\right)^{1/2}=O\left(\left(\pi^{'}\pi\right)^{1/2}N^{1/2}\right).\label{eq:_vZPi}
\end{align}

By Markov's inequality,
\[
J_{212}=\frac{\Delta^{2}v^{'}Z\pi}{\frac{1}{N}\left\Vert \varepsilon\right\Vert ^{2}}=\frac{h^{2}\text{tr}^{2/5}\left(\Sigma^{2}\right)}{N^{2}}O_{\mathbb{P}}\left(\left(\pi^{'}\pi\right)^{1/2}N^{1/2}\right)\overset{p}{\rightarrow}0.
\]

Next, we aim to show the term $\frac{2\Delta^{}X^{'}\varepsilon}{\left\Vert \varepsilon\right\Vert ^{2}}$ converges in probability.
\begin{align*}
J_{22}^{2} & =\frac{4\Delta^{2}\left(Z\pi+v\right)^{'}\varepsilon\varepsilon^{'}\left(Z\pi+v\right)}{\left\Vert \varepsilon\right\Vert ^{4}}\\
 & =\frac{\frac{4}{N^{2}}\Delta^{2}\pi^{'}Z^{'}\varepsilon\varepsilon^{'}Z\pi+\frac{8}{N^{2}}\Delta^{2}\varepsilon^{'}Z\pi+\frac{4}{N^{2}}\Delta^{2}\left(v'\varepsilon\right)^{2}}{\frac{1}{N^{2}}\left\Vert \varepsilon\right\Vert ^{4}}\\
 & :=\frac{J_{221}+J_{222}+J_{223}}{J_{224}}.
\end{align*}
It suffices to show that $J_{221},J_{222},J_{223}\overset{p}{\rightarrow}0$ since $J_{224}=O_{\mathbb{P}}\left(1\right)$ due to Condition (i). Being analogous to Equation \eqref{eq:_vZPi}, we have $\mathbb{E}\left(\pi^{'}Z^{'}\varepsilon\varepsilon^{'}Z\pi\right)=O\left(\pi^{'}\pi\cdot N\right)$. With Markov's inequality,
\[
J_{211}=O_{\mathbb{P}}\left(\frac{h^{2}\text{tr}^{2/5}\left(\Sigma^{2}\right)}{N^{3}}\pi^{'}\pi\cdot N\right)\overset{p}{\rightarrow}0.
\]

Along with Jensen's inequality, we follow the same fashion as in Equation \eqref{eq:_vZPi} to show that $J_{222}\overset{p}{\rightarrow}0.$ By Cauchy-Schwartz's inequality and Condition (i), $J_{223}\leq M\Delta^{2}\left(\frac{1}{N}v^{'}v\right)\left(\frac{1}{N}\varepsilon^{'}\varepsilon\right)=O_{\mathbb{P}}\left(\Delta^{2}\right)\overset{p}{\rightarrow}0.$ We complete the proof of the statement (ii). 
\end{proof}
\begin{proof}[Proof of the statement (i)]
We further expand $J_{1}$ to obtain 
\begin{align*}
J_{1} & =\sqrt{\frac{N^{2}}{2\text{tr}\left(\Sigma^{2}\right)}}\left(u^{'}\underline{S}_{N}u-\frac{1}{N}\text{tr}\left(\underline{S}_{N}\right)\right)+\frac{\Delta^{2}X^{'}ZZ^{'}X}{\left\Vert \varepsilon\right\Vert ^{2}\sqrt{2\text{tr}\left(\Sigma^{2}\right)}}+2\frac{\Delta X^{'}ZZ^{'}\varepsilon}{\left\Vert \varepsilon\right\Vert ^{2}\sqrt{2\text{tr}\left(\Sigma^{2}\right)}}\\
 & -\frac{\text{tr}\left(\underline{S}_{N}\right)}{\sqrt{2\text{tr}\left(\Sigma^{2}\right)}}\left(\frac{\left\Vert X\Delta+\varepsilon\right\Vert ^{2}}{\left\Vert \varepsilon\right\Vert ^{2}}-1\right)\\
 & =:J_{11}+J_{12}+J_{13}-J_{14}.
\end{align*}

It suffices to show that $J_{11}\overset{d}{\rightarrow}\mathcal{N}\left(0,1\right),$ 
\begin{align*}
 & J_{12}+J_{13}-J_{14} \\
 & -\frac{N}{\left\Vert \varepsilon\right\Vert ^{2}\left(2\text{tr}\left(\Sigma^{2}\right)\right)^{1/10}}h^{2}\left(\pi^{'}S_{N}^{2}\pi+\frac{1}{N}v^{'}\underline{S}_{N}v-\frac{1}{N}\text{tr}\left(S_{N}\right)\pi^{'}S_{N}\pi
 - \frac{1}{N^2}tr(S_N)v'v\right)\\
 & +\frac{2 N^{1/2}}{\left\Vert \varepsilon\right\Vert ^{2}\left(2\text{tr}\left(\Sigma^{2}\right)\right)^{3/10}}hv^{'}\left(\frac{1}{N}\text{tr}\left(S_{N}\right)I_{N} - \underline{S}_{N}\right)\varepsilon\\
 & \overset{p}{\rightarrow}0.
\end{align*}
 And, $\left|J_{13}\right|\overset{p}{\rightarrow}0.$ We already show the asymptotic normality of $J_{11}$ in Theorem \ref{thm:_oracle_null}. 

To address $J_{14}$, decompose
\begin{align*}
J_{14} & =\frac{\text{tr}\left(\underline{S}_{N}\right)}{\sqrt{2\text{tr}\left(\Sigma^{2}\right)}}\left(\frac{\Delta^{2}\left(Z\pi+v\right)^{'}\left(Z\pi+v\right)}{\left\Vert \varepsilon\right\Vert ^{2}}+\frac{2\Delta\left(Z\pi+v\right)^{'}\varepsilon}{\left\Vert \varepsilon\right\Vert ^{2}}\right)\\
 & =\frac{\text{tr}\left(\underline{S}_{N}\right)}{\left\Vert \varepsilon\right\Vert ^{2}\sqrt{2\text{tr}\left(\Sigma^{2}\right)}}\left(\Delta^{2}N\pi^{'}S_{N}\pi+2\Delta^{2}v^{'}Z\pi+\Delta^{2}v^{'}v+2\Delta\pi^{'}Z^{'}\varepsilon+2\Delta v^{'}\varepsilon\right)\\
 & =:J_{141}+J_{142}+J_{143}+J_{144}+J_{145},
\end{align*}
where $J_{14i}$'s for $i=1,\ldots,5$, are defined obviously. 
It suffices to show that $J_{142},J_{144}\overset{p}{\rightarrow}0.$ 
Along with Equation \eqref{eq:_vZPi}, we obtain
\begin{align*}
\mathbb{E}\left(\left|J_{142}\right|\right) & =\frac{2\Delta^{2}\text{tr}\left(\underline{S}_{N}\right)}{\left\Vert \varepsilon\right\Vert ^{2}\sqrt{2\text{tr}\left(\Sigma^{2}\right)}}O\left(\left(\pi^{'}\pi\right)^{1/2}N^{1/2}\right)\rightarrow0.
\end{align*}
The limit $J_{142}$ therefore converges in probability thanks to Markov's inequality. 
Being analogous to Equation \eqref{eq:_vZPi}, it follows that $\mathbb{E}\left(\pi^{'}Z^{'}\varepsilon\varepsilon^{'}Z\pi\right)=O\left(\pi^{'}\pi\cdot N\right)$. As a result, we obtain
\[
\mathbb{E}\left(J_{144}^{2}\right)=\frac{4\Delta^{2}\text{tr}^{2}\left(S_{N}\right)}{2\text{tr}\left(\Sigma^{2}\right)\left\Vert \varepsilon\right\Vert ^{4}}\left(\pi^{'}\pi\cdot N\right)\rightarrow0.
\]
We again apply Markov's inequality to imply the convergence in probability of $J_{143}$.

To address $J_{13},$ decompose
\begin{align*}
J_{13} & :=\frac{2\Delta\pi^{'}Z^{'}ZZ^{'}\varepsilon}{\left\Vert \varepsilon\right\Vert ^{2}\sqrt{2\text{tr}\left(\Sigma^{2}\right)}}+\frac{2\Delta v^{'}ZZ^{'}\varepsilon}{\left\Vert \varepsilon\right\Vert ^{2}\sqrt{2\text{tr}\left(\Sigma^{2}\right)}}\\
 & =\frac{2\Delta\pi^{'}Z^{'}ZZ^{'}\varepsilon}{\left\Vert \varepsilon\right\Vert ^{2}\sqrt{2\text{tr}\left(\Sigma^{2}\right)}}+\frac{2\Delta Nv^{'}\underline{S}_{N}\varepsilon}{\left\Vert \varepsilon\right\Vert ^{2}\sqrt{2\text{tr}\left(\Sigma^{2}\right)}}=:J_{131}+J_{132}.
\end{align*}

It suffices to show that $J_{131}$$\overset{p}{\rightarrow}0.$ Recall the expression of $F$, i.e. $F=\sum_{j=1}^{N}f_{j}e_{j}^{'}=F_{j}+f_{j}e_{j}^{'}$ where $F_{j}=\sum_{i\neq j}^{N}f_{i}e_{i}^{'}$ and $Z^{'}=\Sigma^{1/2}F.$ By a direct expansion,
\begin{align*}
J_{131} & =2\frac{\sum_{j=1}^{N}\left(\Delta\pi^{'}\Sigma^{1/2}f_{j}f_{j}^{'}\Sigma\left(F_{j}+f_{j}e_{j}^{'}\right)u\right)}{\left(\frac{1}{T}\left\Vert \varepsilon\right\Vert ^{2}\right)^{1/2}\sqrt{2T\text{tr}\left(\Sigma^{2}\right)}}\\
 & =2\frac{\sum_{j=1}^{N}\left(\Delta\pi^{'}\Sigma^{1/2}f_{j}f_{j}^{'}\Sigma F_{j}u\right)}{\left(\frac{1}{N}\left\Vert \varepsilon\right\Vert ^{2}\right)^{1/2}\sqrt{2N\text{tr}\left(\Sigma^{2}\right)}}+\frac{\sum_{j=1}^{T}\left(\Delta\pi^{'}\Sigma^{1/2}f_{j}f_{j}^{'}\Sigma f_{j}e_{j}^{'}u\right)}{\left(\frac{1}{N}\left\Vert \varepsilon\right\Vert ^{2}\right)^{1/2}\sqrt{2N\text{tr}\left(\Sigma^{2}\right)}}\\
 & =:J_{1311}+J_{1312}.
\end{align*}

\textbf{Addressing $J_{1311}:$} Using the trace property yields,
\begin{align*}
\left(\frac{1}{N}\left\Vert \varepsilon\right\Vert ^{2}\right)^{1/2}J_{1311} & =\frac{1}{\sqrt{2N\text{tr}\left(\Sigma^{2}\right)}}\sum_{j=1}^{N}\left(f_{j}^{'}\Sigma F_{j}u\Delta\pi^{'}\Sigma^{1/2}f_{j}-\Delta\pi^{'}\Sigma^{3/2}F_{j}u\right)\\
 & +\sum_{j=1}^{N}\frac{\Delta\pi^{'}\Sigma^{3/2}F_{j}u}{\sqrt{2N\text{tr}\left(\Sigma^{2}\right)}}:=J_{13111}+J_{13112}.
\end{align*}

By Condition (i) of Theorem \ref{thm:_power_theory}, we know that $\frac{1}{N}\left\Vert \varepsilon\right\Vert ^{2}\overset{p}{\rightarrow}\alpha_{1}<\infty.$ Along with Markov's inequality and Jensen's inequality, it suffices to prove that $\mathbb{E}\left(J_{13111}^{2}\right),\,\mathbb{E}\left(J_{13112}^{2}\right)\rightarrow0.$ By construction, $\sum_{j=1}^{N}F_{j}=\sum_{j=1}^{N}\left(F-f_{j}e_{j}^{'}\right)=\left(N-1\right)F.$ As a result,
\begin{align*}
\mathbb{E}\left(J_{13112}^{2}\right) & =\frac{\mathbb{E}\left(\Delta\pi^{'}\Sigma^{3/2}\sum_{j=1}^{N}F_{j}u\right)^{2}}{2N\text{tr}\left(\Sigma^{2}\right)}\\
 & =\frac{\left(N-1\right)^{2}\mathbb{E}\left(\sum_{j=1}^{N}\Delta\pi^{'}\Sigma^{3/2}f_{j}e_{j}^{'}u\right)^{2}}{2N\text{tr}\left(\Sigma^{2}\right)}\\
 & =\frac{\left(N-1\right)^{2}}{2N\text{tr}\left(\Sigma^{2}\right)}\mathbb{E}\left[\sum_{j=1}^{N}\left(u_{j}^{2}\Delta^{2}\pi^{'}\Sigma^{3/2}f_{j}f_{j}^{'}\Sigma^{3/2}\pi\right)+\sum_{j\neq i}^{N}\left(u_{j}u_{i}\Delta^{2}\pi^{'}\Sigma^{3/2}f_{j}f_{i}^{'}\Sigma^{3/2}\pi\right)\right]\\
 & =\frac{\left(N-1\right)^{2}}{2N\text{tr}\left(\Sigma^{2}\right)}\cdot\Delta^{2}\pi^{'}\Sigma^{3}\pi\cdot\mathbb{E}\left(\sum_{j=1}^{N}u_{j}^{2}\right)\rightarrow0.
\end{align*}

Moreover,
\begin{align*}
\mathbb{E}\left(J_{13111}^{2}\right) & =\frac{1}{2N\text{tr}\left(\Sigma^{2}\right)}\mathbb{E}\left[\sum_{j=1}^{N}\left(f_{j}^{'}\Sigma F_{j}u\Delta\pi^{'}\Sigma^{1/2}f_{j}-\Delta\pi^{'}\Sigma^{3/2}F_{j}u\right)\right]^{2}\\
 & =\frac{1}{2N\text{tr}\left(\Sigma^{2}\right)}\mathbb{E}\left[\sum_{j=1}^{N}\left(f_{j}^{'}\Sigma F_{j}u\Delta\pi^{'}\Sigma^{1/2}f_{j}-\Delta\pi^{'}\Sigma^{3/2}F_{j}u\right)^{2}\right]\\
 & +\frac{1}{2N\text{tr}\left(\Sigma^{2}\right)}\mathbb{E}\left[\sum_{j\neq\ell}^{N}\left(f_{j}^{'}\Sigma F_{j}u\Delta\pi^{'}\Sigma^{1/2}f_{j}-\Delta\pi^{'}\Sigma^{3/2}F_{j}u\right)\left(f_{\ell}^{'}\Sigma F_{j}u\Delta\pi^{'}\Sigma^{1/2}f_{\ell}-\Delta\pi^{'}\Sigma^{3/2}F_{j}u\right)\right]\\
 & =:L_{1}+L_{2}.
\end{align*}
For $\mathbb{E}\left(J_{1312}^{2}\right)\rightarrow0$, it suffices to show $L_{1},L_{2}\rightarrow0.$ Apply Lemma \ref{lem:=000020Concentration=000020inequality} and the trace inequality,
\begin{align*}
L_{1} & =\frac{1}{2N\text{tr}\left(\Sigma^{2}\right)}\mathbb{E}\left[\sum_{j=1}^{N}\mathbb{E}\left(\left.\left(f_{j}^{'}\Sigma F_{j}u\Delta\pi^{'}\Sigma^{1/2}f_{j}-\Delta\pi^{'}\Sigma^{3/2}F_{j}u\right)^{2}\right|\mathcal{F}_{T,*}\right)\right]\\
 & \leq\frac{M\Delta^{2}}{2N\text{tr}\left(\Sigma^{2}\right)}\sum_{j=1}^{N}\mathbb{E}\left[\text{tr}\left(\pi^{'}\Sigma^{3/2}F_{j}uu^{'}F_{j}^{'}\Sigma^{3/2}\pi\right)\right]\\
 & \leq\frac{M\Delta^{2}}{2N\text{tr}\left(\Sigma^{2}\right)}\sum_{j=1}^{N}\mathbb{E}\left[\text{tr}\left(F_{j}^{'}\Sigma^{3/2}\pi\pi^{'}\Sigma^{3/2}F_{j}\right)\right]\\
 & =\frac{MN\left(N-1\right)}{2N\text{tr}\left(\Sigma^{2}\right)}\cdot\pi^{'}\Sigma^{3}\pi,
\end{align*}
where $\mathcal{F}_{T,*}$ is the sigma algebra generated by the product of a sequence $\left\{ f_{i}\right\} _{i=1,i\notin\left\{ j\right\} }^{N}$ and $\left\{ u_{i}\right\} _{i=1}^{N}.$ On the other hand, define $F_{j}=F_{j\ell}+f_{\ell}e_{\ell}^{'},$ 
\begin{align*}
L_{2} & =\frac{1}{2N\text{tr}\left(\Sigma^{2}\right)}\mathbb{E}\left[\sum_{j\neq\ell}^{N}\left(f_{j}^{'}\Sigma F_{j}u\Delta\pi^{'}\Sigma^{1/2}f_{j}-\Delta\pi^{'}\Sigma^{3/2}F_{j}u\right)\left(f_{\ell}^{'}\Sigma F_{j}u\Delta\pi^{'}\Sigma^{1/2}f_{\ell}-\Delta\pi^{'}\Sigma^{3/2}F_{j}u\right)\right]\\
 & =\frac{1}{2N\text{tr}\left(\Sigma^{2}\right)}\mathbb{E}\left[\sum_{j\neq\ell}^{N}\left(f_{j}^{'}\Sigma F_{j\ell}u\Delta\pi^{'}\Sigma^{1/2}f_{j}-\Delta\pi^{'}\Sigma^{3/2}F_{j\ell}u\right)\left(f_{\ell}^{'}\Sigma F_{j}u\Delta\pi^{'}\Sigma^{1/2}f_{\ell}-\Delta\pi^{'}\Sigma^{3/2}F_{j}u\right)\right.\\
 & \left.+\sum_{j\neq\ell}^{N}\left(f_{j}^{'}\Sigma f_{\ell}e_{\ell}^{'}u\Delta\pi^{'}\Sigma^{1/2}f_{j}-\Delta\pi^{'}\Sigma^{3/2}f_{\ell}e_{\ell}^{'}u\right)\left(f_{\ell}^{'}\Sigma F_{j}u\Delta\pi^{'}\Sigma^{1/2}f_{\ell}-\Delta\pi^{'}\Sigma^{3/2}F_{j}u\right)\right]\\
 & =:L_{21}+L_{22}.
\end{align*}

For $L_{2}\rightarrow0$, it suffices to show $L_{21}$ and $L_{22}$ converge. Consider $L_{21},$ given that $\mathcal{F}_{T,**}$ is the sigma algebra generated by the product of a sequence $\left\{ f_{i}\right\} _{i=1,i\notin\left\{ \ell,j\right\} }^{T}$ and $\left\{ u_{i}\right\} _{i=1}^{T}$ ,
\begin{align*}
L_{21} & =\frac{1}{2N\text{tr}\left(\Sigma^{2}\right)}\mathbb{E}\left[\sum_{j\neq\ell}^{N}\mathbb{E}\left(\left(f_{j}^{'}\Sigma F_{j\ell}u\Delta\pi^{'}\Sigma^{1/2}f_{j}-\Delta\pi^{'}\Sigma^{3/2}F_{j\ell}u\right)\right.\right.\\
 & \left.\left.\left.\left(f_{\ell}^{'}\Sigma F_{j}u\Delta\pi^{'}\Sigma^{1/2}f_{\ell}-\Delta\pi^{'}\Sigma^{3/2}F_{j}u\right)\right|\mathcal{F}_{T,**}\right)\right]\\
 & =\frac{1}{2N\text{tr}\left(\Sigma^{2}\right)}\mathbb{E}\left[\sum_{j\neq\ell}^{N}\left(f_{j}\Sigma F_{j\ell}u\Delta\pi^{'}\Sigma^{1/2}f_{j}-\Delta\pi^{'}\Sigma^{3/2}F_{j\ell}u\right)\right.\\
 & \left.\mathbb{E}\left(\left.f_{\ell}^{'}\Sigma F_{j}u\Delta\pi^{'}\Sigma^{1/2}f_{\ell}-\Delta\pi^{'}\Sigma^{3/2}F_{j}u\right|\mathcal{F}_{T,**}\right)\right]=0.
\end{align*}

For $L_{22},$ 
\begin{align*}
L_{22} & =\frac{1}{2N\text{tr}\left(\Sigma^{2}\right)}\mathbb{E}\left[\sum_{j\neq\ell}^{N}u_{\ell}\left(f_{j}^{'}\Sigma f_{\ell}\Delta\pi^{'}\Sigma^{1/2}f_{j}-\Delta\pi^{'}\Sigma^{3/2}f_{\ell}\right)\right.\\
 & \left.\cdot\left(f_{\ell}^{'}\Sigma\left(F_{j\ell}+f_{j}e_{j}^{'}\right)u\Delta\pi^{'}\Sigma^{1/2}f_{\ell}-\Delta\pi^{'}\Sigma^{3/2}\left(F_{j\ell}+f_{j}e_{j}^{'}\right)u\right)\right]\\
 & =\frac{1}{2N\text{tr}\left(\Sigma^{2}\right)}\mathbb{E}\left[\sum_{j\neq\ell}^{N}u_{\ell}\left(f_{j}^{'}\Sigma f_{\ell}\Delta\pi^{'}\Sigma^{1/2}f_{j}-\Delta\pi^{'}\Sigma^{3/2}f_{\ell}\right)\left(f_{\ell}^{'}\Sigma F_{j\ell}u\Delta\pi^{'}\Sigma^{1/2}f_{\ell}-\Delta\pi^{'}\Sigma^{3/2}F_{j\ell}u\right)\right.\\
 & \left.+\sum_{j\neq\ell}^{N}u_{\ell}u_{j}\left(f_{j}^{'}\Sigma f_{\ell}\Delta\pi^{'}\Sigma^{1/2}f_{j}-\Delta\pi^{'}\Sigma^{3/2}f_{\ell}\right)\left(f_{\ell}^{'}\Sigma f_{j}\Delta\pi^{'}\Sigma^{1/2}f_{\ell}-\Delta\pi^{'}\Sigma^{3/2}f_{j}\right)\right]\\
 & =:L_{121}+L_{122}.
\end{align*}

With the same fashion proceeded in $L_{21}$, we address $L_{121}$ by 
\begin{align*}
L_{121} & =\mathbb{E}\left[\sum_{j\neq\ell}^{N}u_{\ell}\mathbb{E}\left(\left.\left(f_{j}^{'}\Sigma f_{\ell}\Delta\pi^{'}\Sigma^{1/2}f_{j}-\Delta\pi^{'}\Sigma^{3/2}f_{\ell}\right)\left(f_{\ell}^{'}\Sigma F_{j\ell}u\Delta\pi^{'}\Sigma^{1/2}f_{\ell}-\Delta\pi^{'}\Sigma^{3/2}F_{j\ell}u\right)\right|\mathcal{F}_{T,*}\right)\right]\\
 & =\mathbb{E}\left[\sum_{j\neq\ell}^{N}u_{\ell}\left(f_{\ell}^{'}\Sigma F_{j\ell}u\Delta\pi^{'}\Sigma^{1/2}f_{\ell}-\Delta\pi^{'}\Sigma^{3/2}F_{j\ell}u\right)\mathbb{E}\left(\left.\left(f_{j}^{'}\Sigma f_{\ell}\Delta\pi^{'}\Sigma^{1/2}f_{j}-\Delta\pi^{'}\Sigma^{3/2}f_{\ell}\right)\right|\mathcal{F}_{T,*}\right)\right]\\
 & =0.
\end{align*}

Let $\tilde{\Sigma}_{k}$ and $\tilde{\Sigma}_{k}^{1/2}$ denote the $k$-th column of matrix $\Sigma$ and $\Sigma^{1/2}$ respectively. Further denote $\left(A\right)_{ij}$ the $\left(i,j\right)$-th entry of a matrix $A$. For $L_{122}$, with Assumption \ref{ass:_errors_iid_X}, it follows that 
\begin{align*}
L_{122} & =\frac{1}{2N\text{tr}\left(\Sigma^{2}\right)}\mathbb{E}\left[\sum_{j\neq\ell}^{N}u_{\ell}u_{j}\left(f_{j}^{'}\Sigma f_{\ell}\Delta\pi^{'}\Sigma^{1/2}f_{j}-\Delta\pi^{'}\Sigma^{3/2}f_{\ell}\right)\left(f_{\ell}^{'}\Sigma f_{j}\Delta\pi^{'}\Sigma^{1/2}f_{\ell}-\Delta\pi^{'}\Sigma^{3/2}f_{j}\right)\right]\\
 & =\frac{\Delta^{2}}{2N\text{tr}\left(\Sigma^{2}\right)}\mathbb{E}\left[\sum_{j\neq\ell}^{N}u_{\ell}u_{j}\cdot\mathbb{E}\left(\left.f_{j}^{'}\Sigma f_{\ell}f_{\ell}^{'}\Sigma f_{j}\cdot f_{\ell}^{'}\Sigma^{1/2}\pi\pi^{'}\Sigma^{1/2}f_{j}\right|f_{\ell},u_{1},\ldots,u_{T}\right)\right]\\
 & =\frac{\Delta^{2}\mathbb{E}\left(f_{11}^{3}\right)}{2N\text{tr}\left(\Sigma^{2}\right)}\mathbb{E}\left[\sum_{j\neq\ell}^{N}u_{\ell}u_{j}\cdot\sum_{k=1}^{p}\left(f_{\ell}^{'}\tilde{\Sigma}_{k}\tilde{\Sigma}_{k}^{'}f_{\ell}\right)\left(\tilde{\Sigma}_{k}^{1/2}\pi\pi^{'}\Sigma^{1/2}f_{\ell}\right)\right]\\
 & =\frac{\Delta^{2}\mathbb{E}\left(f_{11}^{3}\right)}{2N\text{tr}\left(\Sigma^{2}\right)}\mathbb{E}\left[\sum_{j\neq\ell}^{N}u_{\ell}u_{j}\cdot\sum_{k=1}^{p}\sum_{s=1}^{p}\left(\tilde{\Sigma}_{k}\tilde{\Sigma}_{k}^{'}\right)_{ss}\left(\pi^{'}\tilde{\Sigma}_{s}^{1/2}\tilde{\Sigma}_{k}^{1/2}\pi\right)\right].
\end{align*}

Using the majorization property that $\underset{ss\in\left\{ 1,\ldots,p\right\} }{\text{max}}\left(\tilde{\Sigma}_{k}\tilde{\Sigma}_{k}^{'}\right)_{ss}\leq\lambda_{\max}\left(\Sigma\right),$ $\sum_{s=1}^{p}\pi^{'}\tilde{\Sigma}_{s}^{1/2}=\pi^{'}\Sigma^{1/2}\mathds{1}_{p}$ and Cauchy-Schwarz's inequality, it follows that 
\begin{align*}
0\leq\left|L_{122}\right| & \leq\frac{M\Delta^{2}}{2N\text{tr}\left(\Sigma^{2}\right)}\lambda_{\max}\left(\Sigma\right)\left(\pi^{'}\Sigma^{1/2}\mathds{1}_{p}\right)^{2}\left|\mathbb{E}\left(\sum_{j\neq\ell}^{N}u_{\ell}u_{j}\right)\right|\\
 & \leq\frac{M\Delta^{2}}{2N}\cdot\frac{\lambda_{\max}\left(\Sigma^{2}\right)}{\frac{1}{p}\text{tr}\left(\Sigma^{2}\right)}\cdot\pi^{'}\pi\cdot\mathbb{E}\left(\sum_{j=1}^{N}u_{j}\right)^{2}\\
 & \leq\frac{M\Delta^{2}}{2N}\cdot\frac{\lambda_{\max}\left(\Sigma^{2}\right)}{\frac{1}{p}\text{tr}\left(\Sigma^{2}\right)}\cdot\pi^{'}\pi\cdot N\mathbb{E}\left(\sum_{j=1}^{N}u_{j}^{2}\right)\rightarrow0.
\end{align*}

\textbf{Addressing $J_{1312}:$} We rewrite
\begin{align*}
\left(\frac{1}{T}\left\Vert \varepsilon\right\Vert ^{2}\right)^{1/2}J_{1312} & =\frac{\sum_{j=1}^{N}\left(\Delta\pi^{'}\Sigma^{1/2}f_{j}f_{j}^{'}\Sigma f_{j}e_{j}^{'}u\right)}{\sqrt{2N\text{tr}\left(\Sigma^{2}\right)}}\\
 & =\frac{1}{\sqrt{2N\text{tr}\left(\Sigma^{2}\right)}}\left[\sum_{j=1}^{N}\left(f_{j}^{'}\Sigma f_{j}e_{j}^{'}u\Delta\pi^{'}\Sigma^{1/2}f_{j}-\mathbb{E}\left(\left.f_{j}^{'}\Sigma f_{j}e_{j}^{'}u\Delta\pi^{'}\Sigma^{1/2}f_{j}\right|u\right)\right)\right]\\
 & +\sum_{j=1}^{N}\frac{u_{j}\mathbb{E}\left(\left.f_{j}^{'}\Sigma f_{j}\Delta^{'}\Sigma^{1/2}f_{j}\right|u\right)}{\sqrt{2N\text{tr}\left(\Sigma^{2}\right)}}=:J_{13121}+J_{13122}.
\end{align*}

It suffices to show that $J_{13121}$ and $J_{13122}$ converge in probability. To address $J_{13122},$ we apply Cauchy-Schwartz's inequality
\begin{align*}
0\leq J_{13122}^{2} & =\frac{\left(\sum_{j=1}^{N}u_{j}\mathbb{E}\left(\left.f_{j}^{'}\Sigma f_{j}\Delta\pi^{'}\Sigma^{1/2}f_{j}\right|u\right)\right)^{2}}{2N\text{tr}\left(\Sigma^{2}\right)}\\
 & =\frac{N\left(\mathbb{E}\left(f_{1}^{'}\Sigma f_{1}\Delta\pi^{'}\Sigma^{1/2}f_{1}\right)\right)^{2}\sum_{j=1}^{N}u_{j}^{2}}{2N\text{tr}\left(\Sigma^{2}\right)}\\
 & =\frac{\left(\sum_{j=1}^{p}\Sigma_{jj}\Delta\pi^{'}\tilde{\Sigma_{j}}^{1/2}\mathbb{E}\left(f_{11}^{3}\right)\right)^{2}}{2\text{tr}\left(\Sigma^{2}\right)}\\
 & \leq\frac{M\lambda_{\max}\left(\Sigma\right)}{\text{tr}\left(\Sigma^{2}\right)}\left(\Delta\pi^{'}\Sigma^{1/2}\mathds{1}_{p}\right)^{2}\\
 & \leq\frac{M\lambda_{\max}\left(\Sigma^{2}\right)}{\frac{1}{K}\text{tr}\left(\Sigma^{2}\right)}\Delta^{2}\pi^{'}\pi\rightarrow0.
\end{align*}

The argument holds thanks to Markov's inequality. With the same fashion, we address $J_{13121}$ by showing that 
\begin{align*}
\mathbb{E}\left|J_{13121}\right|^{2} & =\frac{1}{2N\text{tr}\left(\Sigma^{2}\right)}\mathbb{E}\left[\sum_{j=1}^{N}\left(f_{j}^{'}\Sigma f_{j}e_{j}^{'}u\Delta\pi^{'}\Sigma^{1/2}f_{j}-\mathbb{E}\left(\left.f_{j}^{'}\Sigma f_{j}e_{j}^{'}u\Delta\pi^{'}\Sigma^{1/2}f_{j}\right|u\right)\right)\right]^{2}\\
 & =\frac{1}{2N\text{tr}\left(\Sigma^{2}\right)}\mathbb{E}\left[\sum_{j=1}^{N}u_{j}^{2}\left(f_{j}^{'}\Sigma f_{j}\Delta\pi^{'}\Sigma^{1/2}f_{j}-\mathbb{E}\left(f_{j}^{'}\Sigma f_{j}\Delta\pi^{'}\Sigma^{1/2}f_{j}\right)\right)^{2}+\sum_{j\neq i}^{N}u_{i}u_{j}\cdot\right.\\
 & \left.\left(f_{j}^{'}\Sigma f_{j}\Delta\pi^{'}\Sigma^{1/2}f_{j}-\mathbb{E}\left(f_{j}^{'}\Sigma f_{j}\Delta\pi^{'}\Sigma^{1/2}f_{j}\right)\right)\left(f_{i}^{'}\Sigma f_{i}\Delta\pi^{'}\Sigma^{1/2}f_{i}-\mathbb{E}\left(f_{i}^{'}\Sigma f_{i}\Delta\pi^{'}\Sigma^{1/2}f_{i}\right)\right)\right]\\
 & =:L_{1}+L_{2}.
\end{align*}

It suffices to show that $L_{1},L_{2}\rightarrow0$. For $L_{1},$ we further decompose
\begin{align*}
L_{1} & =\frac{1}{2N\text{tr}\left(\Sigma^{2}\right)}\mathbb{E}\left(\sum_{j=1}^{T}u_{j}^{2}\mathbb{E}\left[\left(f_{j}^{'}\Sigma f_{j}-\text{tr}\left(\Sigma\right)\right)\Delta\pi^{'}\Sigma^{1/2}f_{j}\right.\right.\\
 & \left.\left.-\mathbb{E}\left(f_{j}^{'}\Sigma f_{j}\Delta\pi^{'}\Sigma^{1/2}f_{j}\right)+\text{tr}\left(\Sigma\right)\Delta\pi^{'}\Sigma^{1/2}f_{j}\right]^{2}\right)\\
 & \leq\frac{1}{2N\text{tr}\left(\Sigma^{2}\right)}\cdot\mathbb{E}\left(1\right)\cdot\left(\mathbb{E}\left[\left(f_{j}^{'}\Sigma f_{j}-\text{tr}\left(\Sigma\right)\right)\Delta\pi^{'}\Sigma^{1/2}f_{j}\right]^{2}\right.\\
 & \left.+\mathbb{E}^{2}\left(f_{j}^{'}\Sigma f_{j}\Delta\pi^{'}\Sigma^{1/2}f_{j}\right)+\text{tr}^{2}\left(\Sigma\right)\mathbb{E}\left(\Delta\pi^{'}\Sigma^{1/2}f_{j}\right)^{2}\right)\\
 & \leq\frac{1}{2N\text{tr}\left(\Sigma^{2}\right)}\left(\mathbb{E}\left[\left(f_{j}^{'}\Sigma f_{j}-\text{tr}\left(\Sigma\right)\right)^{2}\left(\Delta\pi^{'}\Sigma^{1/2}f_{j}\right)^{2}\right]\right)\\
 & +\frac{\lambda_{\max}\left(\Sigma^{2}\right)\left(\Delta\pi^{'}\Sigma\mathds{1}\right)^{2}}{2N\text{tr}\left(\Sigma^{2}\right)}+\frac{\text{tr}^{2}\left(\Sigma\right)\Delta^{2}\pi^{'}\Sigma\pi}{2N\text{tr}\left(\Sigma^{2}\right)}\\
 & =:L_{11}+L_{12}+L_{13}.
\end{align*}
 Applying Holder's inequality to $L_{11}$ yields,
\begin{align*}
0\leq L_{11} & \leq\frac{1}{2N\text{tr}\left(\Sigma^{2}\right)}\left(\left[\mathbb{E}\left(f_{j}^{'}\Sigma f_{j}-\text{tr}\left(\Sigma\right)\right)^{4}\mathbb{E}\left(\Delta\pi^{'}\Sigma^{1/2}f_{j}\right)^{4}\right]^{1/2}\right)\\
 & =\frac{1}{2N\text{tr}\left(\Sigma^{2}\right)}\cdot\mathbb{E}^{1/2}\left(f_{j}^{'}\Sigma f_{j}-\text{tr}\left(\Sigma\right)\right)^{4}\cdot\mathbb{E}^{1/2}\left(\Delta f_{j}^{'}\Sigma^{1/2}\pi\pi^{'}\Sigma^{1/2}f_{j}\right)^{2}\\
 & \leq\frac{K\Delta^{2}}{2N\text{tr}\left(\Sigma^{2}\right)}\cdot\text{tr}\left(\Sigma^{2}\right)\cdot\pi^{'}\Sigma\pi\rightarrow0.
\end{align*}

The last inequality holds due to Condition (ii), and Lemma \ref{lem:=000020Concentration=000020inequality}. With the same fashion, $L_{12}$ converges as $0\leq L_{12}\leq\frac{K\Delta^{2}\lambda_{\max}^{2}\left(\Sigma^{2}\right)}{\text{tr}\left(\Sigma^{2}\right)}\pi^{'}\pi\rightarrow0$. And, $0\leq L_{13}\leq\frac{\Delta^{2}\text{tr}^{2}\left(\Sigma\right)\lambda_{\max}\left(\Sigma^{2}\right)}{N\text{tr}\left(\Sigma^{2}\right)}\pi^{'}\pi\rightarrow0.$ 

For $J_{12},$ by definition, we directly have 
\begin{align*}
J_{12} & =\frac{\Delta^{2}\left(Z\pi+v\right)^{'}ZZ^{'}\left(Z\pi+v\right)}{\left\Vert \varepsilon\right\Vert ^{2}\sqrt{2\text{tr}\left(\Sigma^{2}\right)}}\\
 & =\frac{\Delta^{2}N^{2}\pi^{'}S_{N}^{2}\pi+2\Delta^{2}v^{'}ZZ^{'}Z\pi+\Delta^{2}Nv^{'}\underline{S}_{N}v}{\left\Vert \varepsilon\right\Vert ^{2}\sqrt{2\text{tr}\left(\Sigma^{2}\right)}}\\
 & :=J_{121}+J_{122}+J_{123},
\end{align*}
where $J_{121},J_{122}$ and $J_{123}$ are defined obviously. It suffices to show that $J_{122}\overset{p}{\rightarrow}0.$ By applying the same proof techniques used for $J_{131}$, we conclude that $J_{122}$ converges in probability. This completes the proof.

\end{proof}

\end{document}